\documentclass[longauth, traditabstract]{aa} 
\usepackage{graphicx}
\usepackage{epstopdf}
\usepackage{txfonts}
\usepackage{natbib}
\begin{document}

\newcommand{\water}{H$_2$O}
\newcommand{\iwater}{H$_2^{\;18}$O}
\newcommand{\iiwater}{H$_2^{\;17}$O}
\newcommand{\asec}{$^{\prime\prime}$}
\newcommand{\vlsr}{v$_{\rm LSR}$}
\newcommand{\kms}{$\:$km~s$^{-1}$}
\newcommand{\tr}[6]{#1$_{#2 #3}\:-\:$#4$_{#5 #6}$}
\newcommand{\etal}{et$\:$al.}

\setcounter{bottomnumber}{5}
\setcounter{totalnumber}{10}
\renewcommand{\bottomfraction}{1.0}
\renewcommand{\textfraction}{0.0}
\renewcommand{\dblfloatpagefraction}{1.00}

\title{{\em Herschel} observations of EXtra-Ordinary Sources (HEXOS): Observations of
  H$_2$O and its isotopologues towards Orion~KL\thanks{Herschel is an ESA space
    observatory with science instruments provided by European-led
    Principal Investigator consortia and with important participation
    from NASA.}}

\author{
G.~J.~Melnick,\inst{1}
V.~Tolls,\inst{1}
D.~A.~Neufeld,\inst{2}
E.~A.~Bergin,\inst{3}
T.~G.~Phillips,\inst{4}
S.~Wang,\inst{3}
N.~R.~Crockett,\inst{3}
T.~A.~Bell,\inst{4}
G.A.~Blake\inst{5}
S.~Cabrit,\inst{23}
E.~Caux,\inst{6,7}
C.~Ceccarelli,\inst{8}
J.~Cernicharo,\inst{9}
C.~Comito,\inst{10}
F.~Daniel,\inst{9,11}
M.-L.~Dubernet,\inst{12,13}
M.~Emprechtinger,\inst{4}
P.~Encrenaz,\inst{11}
E.~Falgarone,\inst{11}
M.~Gerin,\inst{11}
T.~F.~Giesen,\inst{14}
J.~R.~Goicoechea,\inst{9}
P.~F.~Goldsmith,\inst{15}
E.~Herbst,\inst{16}
C.~Joblin,\inst{4,5}
D.~Johnstone,\inst{17}
W.~D.  Langer\inst{15}
W.D.~Latter\inst{18} 
D.~C.~Lis,\inst{4}
S.~D.~Lord,\inst{18}
S.~Maret,\inst{8}
P.~G.~Martin,\inst{19}
K.~M.~Menten,\inst{10}
P.~Morris,\inst{18}
H.~S.~P. M\"uller,\inst{14}
J.~A.~Murphy,\inst{20}
V.~Ossenkopf,\inst{14,21}
L.~Pagani,\inst{23}
J.~C.~Pearson,\inst{15}
M.~P\'erault,\inst{11}
R.~Plume,\inst{22}
S.-L.~Qin,\inst{14}
M.~Salez,\inst{23}
P.~Schilke,\inst{10,14}
S.~Schlemmer,\inst{14}
J.~Stutzki,\inst{14}
N.~Trappe,\inst{20}
F.~F.~S.~van der Tak,\inst{21}
C.~Vastel,\inst{6,7}
H.~W.~Yorke,\inst{15}
S.~Yu,\inst{15}
\and
J.~Zmuidzinas,\inst{4}
}
\institute {Harvard-Smithsonian Center for Astrophysics (CfA), 60 Garden Street, 
Cambridge MA 02138, USA\\
             \email{gmelnick@cfa.harvard.edu}
\and  Department of Physics and Astronomy, Johns Hopkins University, 3400 North Charles Street, Baltimore, MD 21218, USA
\and Department of Astronomy, University of Michigan, 500 Church Street, Ann Arbor, MI 48109, USA 
\and California Institute of Technology, Cahill Center for Astronomy and Astrophysics 301-17, Pasadena, CA 91125 USA
\and California Institute of Technology, Division of Geological and Planetary Sciences, MS 150-21, Pasadena, CA 91125, USA
\and Centre d'\'etude Spatiale des Rayonnements, Universit\'e de Toulouse [UPS], 31062 Toulouse Cedex 9, France
\and CNRS/INSU, UMR 5187, 9 avenue du Colonel Roche, 31028 Toulouse Cedex 4, France
\and Laboratoire d'Astrophysique de l'Observatoire de Grenoble, 
BP 53, 38041 Grenoble, Cedex 9, France.
\and Centro de Astrobiolog\'ia (CSIC/INTA), Laboratiorio de Astrof\'isica Molecular, Ctra. de Torrej\'on a Ajalvir, km 4
28850, Torrej\'on de Ardoz, Madrid, Spain
\and Max-Planck-Institut f\"ur Radioastronomie, Auf dem H\"ugel 69, 53121 Bonn, Germany 
\and LERMA, CNRS UMR8112, Observatoire de Paris and \'Ecole Normale Sup\'erieure, 24 Rue Lhomond, 75231 Paris Cedex 05, France
\and LPMAA, UMR7092, Universit\'e Pierre et Marie Curie,  Paris, France
\and  LUTH, UMR8102, Observatoire de Paris, Meudon, France
\and I. Physikalisches Institut, Universit\"at zu K\"oln,
              Z\"ulpicher Str. 77, 50937 K\"oln, Germany
\and Jet Propulsion Laboratory,  Caltech, Pasadena, CA 91109, USA
\and Departments of Physics, Astronomy and Chemistry, Ohio State University, Columbus, OH 43210, USA
\and National Research Council Canada, Herzberg Institute of Astrophysics, 5071 West Saanich Road, Victoria, BC V9E 2E7, Canada 
\and Infrared Processing and Analysis Center, California Institute of Technology, MS 100-22, Pasadena, CA 91125
\and Canadian Institute for Theoretical Astrophysics, University of Toronto, 60 St George St, Toronto, ON M5S 3H8, Canada
\and  National University of Ireland Maynooth. Ireland
\and SRON Netherlands Institute for Space Research, PO Box 800, 9700 AV, Groningen, The Netherlands
\and Department of Physics and Astronomy, University of Calgary, 2500
University Drive NW, Calgary, AB T2N 1N4, Canada
\and LERMA \& UMR8112 du CNRS, Observatoire de
 Paris, 61, Av. de l'Observatoire, 75014 Paris, France
}


\abstract{We report the detection of more than 48 velocity-resolved ground rotational state 
transitions of H$_2^{\;16}$O, \iwater, and \iiwater\ -- most for the first time -- 
in both emission and absorption toward Orion~KL using {\em Herschel}/HIFI.  We show that a simple
fit, constrained to match the known emission and absorption components along the
line of sight, is in excellent agreement with the spectral profiles 
of all the water lines.  Using the measured \iwater\ line fluxes, which are less
affected by line opacity than their H$_2^{\;16}$O counterparts, and an escape probability method,
the column densities of \iwater\ associated with each emission component are derived. 
We infer total water abundances of
7.4$\times$10$^{-5}$, 1.0$\times$10$^{-5}$, and 1.6$\times$10$^{-5}$ for the
plateau, hot core, and extended warm gas, respectively.  In the case of the plateau, this value 
is consistent with previous measures of the Orion-KL water abundance 
as well as those of other molecular outflows.  In the case of the hot core and extended 
warm gas, these values are somewhat higher than water 
abundances derived for other quiescent clouds, suggesting that these
regions are likely experiencing enhanced water-ice sublimation from (and reduced
freeze-out onto) grain surfaces due to the warmer dust in these sources.}

   \titlerunning{Water in Orion}
	\authorrunning{Melnick et al.}
   \maketitle
%

\section{Introduction}

During its 6-year mission, the Submillimeter Wave Astronomy Satellite (SWAS)
surveyed more than 300 galactic sources and more than 6800 lines-of-sight
(Melnick \etal~2000a), yet none produced
stronger water emission than the line of sight toward Orion-KL.  The source of this emission
was attributed primarily to the chemistry and excitation accompanying the exceptionally
powerful outflows emanating from the BN/KL region (Harwit \etal~1998; Wright \etal~2000;
Melnick \etal~2000b; Cernicharo \etal~2006; Lerate \etal~2006); 
however, many sources possessing 
physical conditions favorable to the production of strong water emission -- e.g. high densities
and temperatures -- are known to exist close to KL and could very likely be significant
contributors to the water emission detected by ISO, SWAS, and {\em Odin}.  
Unfortunately, with access to only the
ground-state 1$_{10\,}-\,$1$_{01}$ transition of ortho-H$_2^{\;16}$O \footnote{Also 
referred to simply as \water} and \iwater, even the velocity-resolved SWAS and {\em Odin}
measurements were limited in what could be inferred
about the various components giving rise to the strong water emission.

The availability of the {\em Herschel}/HIFI instrument (deGraauw \etal~2010)
with its extended frequency coverage and higher angular resolution, now
permits a more detailed examination of the conditions responsible for the water emission
toward Orion-KL.  Here we report the detection of 21 \water, 15 \iwater, and 12 \iiwater\
velocity-resolved lines toward this source obtained as part of the HEXOS program 
(Bergin \etal~2010).

In this paper, we present an analysis of the sources of the water emission based
upon the lower-opacity lines of \iwater.  We also show that the approach taken in this analysis
holds great promise when applied to the \water\ and \iiwater\ lines, which will be
pursued in a future paper.

\section{Observations and results}

The HIFI observations presented here were carried out in March and April 2010
using the spectral scan dual beam switch (DBS) mode pointed towards
Orion-KL $\alpha_{J2000} = 5^h35^m14.3^s$ and $\delta_{J2000} =
-5^{\circ}22'33.7''$.
All observations were obtained with a beamsize of $\sim\,$(22$\,/\,\nu_{\rm THz})''$
and reference beams approximately
3$^{\prime}$ east and west, which is roughly orthogonal to the orientation 
of the Orion molecular ridge (e.g., Ungerechts \etal~1997).
However, water emission 
is extended in Orion (Snell \etal~2000) and the reference beam may contain 
some contamination from a narrow ($\Delta$v$\,\sim 3-5$ km s$^{-1}$) component 
centered at $\sim\,$9 km s$^{-1}$.   We utilized the wide band spectrometer 
providing a spectral resolution of 1.1 MHz over a 4~GHz IF bandwidth.
The data presented here are from a range of HIFI bands obtained as part of the 
HEXOS program.  These data were reduced and converted to single side band as 
described by (Bergin \etal~2010), with additional analysis performed
at the CfA.  In our study, we adopt a uniform 
main beam efficiency of 70\%.

Because of
flux differences between the H- and the V-polarizations, which are most likely due to the
known pointing offset between the two beams, we use only the H-polarization data for 
our analysis.  The spectra for all H$_2$O, \iiwater, and \iwater\ lines were extracted 
from the more extended HEXOS spectral scan data using the JPL Spectral Line Catalog 
(Pickett \etal~1998) for identification. 
Finally, the continuum offset appropriate to each line was determined directly from 
emission-free spectral regions near each line.

\setcounter{figure}{0}


Figures~1 and 2 show the spectra of \iwater\ and  \water\ plus
\iiwater, respectively.  These spectra
span a broad range of excitation conditions, ranging in upper-level energies between
53~K and more than 1000~K.  
All spectra have been examined for severe blending using the CLASS-Weeds tool 
(Maret \etal~2010), the JPL Spectral Line Catalog, or visual evidence of
non-smooth water line wings.  Blended lines were excluded from the following 
analysis. 

\section{Analysis}

The goals of the present effort are twofold: (1) isolate the components giving rise 
to the water  emission we detect; and, (2) model these components in a way that
best reproduces the measured line fluxes and profiles.  To do this, we focus
here on the observed \iwater\ lines.  These lines have been detected over a broad
range of excitation conditions with high signal-to-noise ratios and are much less 
affected by optical depth effects than their H$_2^{\;16}$O counterparts, making the
analysis more straightforward.  In addition, the $^{16}$O:$^{18}$O ratio is
well known (i.e., $\sim\,$500) and not believed to vary significantly between sources,
making the conversion from inferred \iwater\ abundance to H$_2^{\;16}$O abundance 
robust.

Step 1 -- isolate the components:  The lines exhibit complex profiles which we attribute
to a combination 
of emission and absorption components along the line of sight.  To isolate what we 
believe are the three predominant emission components within the HIFI beams --
namely the plateau molecular outflow, the hot core, and an extended region of
gas composed of the compact ridge plus the warmer, denser portion of the extended
ridge near KL (cf. Blake \etal~1987) -- we adopt a line-fitting 
strategy that
fixes the well-established characteristics of these regions, such as their v$_{\rm LSR}$,
and, in some cases, the typical line width, and leaves as free fitting parameters such
quantities as the line strengths.  

In addition to the three emission components,
we include the effects of absorption by foreground material in two distinct kinematic 
components: a narrow component near 7 km~s$^{-1}$, and a broad 
component centered at an LSR velocity of $-$5.1 km~s$^{-1}$.  While the presence 
of these absorption components is clearly required to fit the observed water line profiles, 
particularly in the case of low-lying transitions of H$_2^{\;16}$O, the existence 
of foreground absorbing material at these velocities has been independently confirmed 
by HIFI observations of HF (Phillips \etal~2010), OH$^+$ and 
\water$^+$ (Gupta \etal~2010), as well as CRIRES observations of the 
fundamental CO vibrational band (Beuther \etal~2010).
The narrow component arises in quiescent gas, while the 
broad, blueshifted component represents outflowing material, presumably associated 
primarily with the Low Velocity Flow (Genzel \& Stutzki 1989).   
For the lower-lying transitions, these absorption components account for pronounced 
asymmetries in the line shapes, as well as the absorption feature close to the systemic source velocity 
(although we note here that narrow line emission in the reference beam is potentially 
a contributor to this absorption feature observed in the very lowest transitions.)  Even in the case of 
\iwater, transitions to the ground states of ortho- or para-\iwater\ (i.e., 2$_{12\:}-\,$1$_{01}$,
1$_{10\:}-\,$1$_{01}$ and 1$_{11\:}-\,$0$_{00}$) are affected by foreground absorption.
Indeed, in the 1$_{11\:}-\,$0$_{00}$ and 2$_{12\:}-\,$1$_{01}$ \iwater\ transitions, 
where the continuum brightness temperature is greatest, the blueshifted absorption 
feature can cause the observed antenna temperature to dip below the continuum level.

\begin{table}[tdp]
\caption{Fixed and varied parameters in water-line fits}
\vspace{-2.6mm}
\begin{center}
\begin{tabular}{lccc} \hline
\rule{0mm}{4.1mm} & ~Peak $T_A^*$~ & \vlsr\ & ~FWHM~ \\*[0.1mm]
\hspace{4mm}Source\hspace{24mm} & (K) & ~(km s$^{-1}$)~ & (km s$^{-1}$) \\*[1mm] \hline
\rule{0mm}{5mm}Plateau \dotfill\ & Varied & $+$6.9~ & Varied \\*[0.7mm]
Hot Core \dotfill\ & Varied & $+$5.2~ & 10.0 \\*[0.7mm]
Extended warm gas \dotfill\ & Varied & ~~$+$8.25~ & ~~2$\,-\,$8\raisebox{0.7mm}{$\,\dagger$} \\*[0.7mm]
Narrow absorption \dotfill\ & Varied & ~~$+$6.88~ & 6.70 \\*[0.7mm]
Broad absorption \dotfill\ & Varied & $-$5.1~~~ & 30.0 \\*[0.7mm] \hline
\end{tabular}
\end{center}
\vspace{-4.9mm}
\phantom{0}\hspace{2mm}\raisebox{1mm}{$\dagger$}~The FWHM was constrained to 
vary between only 2 and 8\kms.
\vspace{-2.8mm}
\label{default}
\end{table}%

\begin{table*}[t]
\begin{center}
\caption{Best-fit radiative transfer model parameters for Orion \iwater\ emission components}
\vspace{-2.9mm}
\begin{tabular}{lccccccc} \hline
\rule{0mm}{4.5mm}  & $\Delta$v & $T_{\rm gas}$ & $n({\rm H_2})$ & 
   $T_{\rm dust}$\raisebox{0.7mm}{$\dagger$} &   & $N($ortho/para-\iwater) & 
   ~\raisebox{-1mm}{Total Inferred}~ \\*[1mm]
\phantom{0}\hspace{4mm}Source\hspace{22.1mm}  & (km s$^{-1}$) & 
   (K) & (cm$^{-3}$) & (K) & $\theta_{\rm source}$ & (cm$^{-2}$) & 
   \water\ Abundance\raisebox{0.7mm}{$\,\ddagger$} \\*[1.2mm] \hline
\rule{0mm}{5mm}Extended warm gas \dotfill\ &  2$\,-\,$8 & ~~75 & 2$\times$10$^6$ & 30 &
   20\asec\ & ~~7.4$\times$10$^{15}$~(ortho)~/~2.0$\times$10$^{15}$~(para)~~ 
   & 1.6$\times$10$^{-5}$ \\*[1.2mm]
Plateau \dotfill\ &  ~20$\,-\,$34~ & ~~188~~ & ~~2$\times$10$^6$~~ & 113.6 & 24\asec\ &
   1.2$\times$10$^{16}$~(ortho)~/~2.8$\times$10$^{15}$~(para) & 7.4$\times$10$^{-5}$ \\*[1.2mm]
Hot core \dotfill\  & 10 & ~150~ & 1$\times$10$^7$ & ~180~ & ~~5\asec\ & 
   1.5$\times$10$^{16}$~(ortho)~/~5.0$\times$10$^{15}$~(para) & 1.0$\times$10$^{-5}$ \\*[1.1mm] \hline
\end{tabular}
\end{center}
\vspace{-2.4mm}
\raisebox{0.7mm}{$\dagger\,$}Greybody fit to the Orion continuum of the form:~
$B_{\tilde{\nu\,}}(T_{\rm dust}) \times\ (0.0233\,\tilde{\nu})^{0.486}$, where $\tilde{\nu}$ 
is wavelength in wavenumbers. 
\raisebox{0.7mm}{$\ddagger\,$}Assumes $^{16}$O/$^{18}$O$\,=\,$500 \\ \phantom{0.}and
$N$(H$_2)\,=\:$3$\times$10$^{23}$,
1$\times$10$^{23}$, and 1$\times$10$^{24}$ cm$^{-2}$ in a 30\asec\ beam for
the extended ridge, plateau, and
hot core, respectively (Blake \etal~1987).
\vspace{-1.5mm}
\label{default}
\end{table*}%

Thus, fits to all lines were made using the expression:

\vspace{-3.8mm}

\begin{eqnarray}
{\rm Fitted~Line}~(T_A^{\:*}) & = & {\rm\left(Continuum~Offset + G_{plat} + G_{hc} + G_{ewg}\right)}
     \nonumber \\*[0.5mm]
   &  & \phantom{0}\times\;\,{\rm Exp\left[ - \left(G_{na} + G_{ba}\right)\right]}\;,
\end{eqnarray}

\vspace{-1mm}

\noindent where $\rm G_{plat}$, $\rm G_{hc}$, $\rm G_{ewg}$, $\rm G_{na}$ and
$\rm G_{ba}$ are Gaussian components representing the plateau, hot core, extended warm gas,
narrow absorbing feature, and broad absorbing feature, respectively.  
Table~1 provides the fit
parameters fixed by previous measurements and those that were allowed to vary,
unconstrained, in order to 
obtain the best fit to the line profiles.

Step 2 -- model the \iwater\ emission components: The results of Step 1 are a set of 
best-fit integrated intensities for each component and transition, including 
the absorption features, that sum to reproduce the line flux
and profile for each ortho- and para \iwater\ line.  
In this paper, we focus on the
emission components only; analysis of the physical conditions
associated with the absorption components will be undertaken following the results of
a soon-to-be-completed water map toward Orion-KL.
To assess how the \iwater\ line strengths
constrain the water abundance in each component, the equilibrium level
populations of all \water\ ortho and para rotational levels of the ground
vibrational state with energies {\em E/k} up to 2000~K have been calculated using 
an escape probability method that includes the necessary effects of radiative excitations due to
dust emission embedded within each component.  
It is assumed that the water molecules see 4$\pi$ steradians of dust emission from within
each component.
The velocity gradient for each transition is assumed to be equal to $\Delta$v$\,n$(H$_2$)/$N$(H$_2$),
where the line width, $\Delta$v, for each line for each component is taken from the best fit in Step 1,
and $n$(H$_2$) and
$N$(H$_2$) are the volume and column densities of H$_2$, respectively.
The rate coefficients for collisions 
between ortho- and para-H$_2$ and ortho- and para-\water\ calculated by 
Faure \etal~(2007) are used, and the H$_2$ ortho-to-para ratio is assumed to
be the LTE value at the gas temperature of each component.  Finally, the calculations incorporate the 
beam size and aperture efficiency appropriate to each transition.

More than 90\% of the presently observed \water\ total line flux (and $>\,$98\% of the \iwater\ 
and \iiwater\ total line flux) 
lies in transitions with $E_{\rm upper}\leq\:$600~K.  Thus, we focus our modeling
efforts primarily on reproducing the flux and profiles for these transitions.
The H$_2$ density, gas and dust temperatures, source size,
and ortho- and para-\iwater\ column densities were varied to best match the
inferred line fluxes for each emission component.  The values yielding the best
fit to the data are provided in Table~2.  The line profiles resulting from the 
radiative transfer model calculations for the emission components and 
Step~1 line-fits to the absorption components
are shown as the red curves superposed 
on the observed spectra in Fig.~1.  The models summarized in Table~2 provide
a remarkably good match to the data, though the deviation between the models
and the observed spectra for the higher-energy \iwater\ transitions clearly illustrates the
shortcomings of single-value models for each component as small amounts of
hotter gas are not accounted for.

\begin{figure}
   \centering
   \vspace{1.1mm}
   \includegraphics[width=0.80\columnwidth]{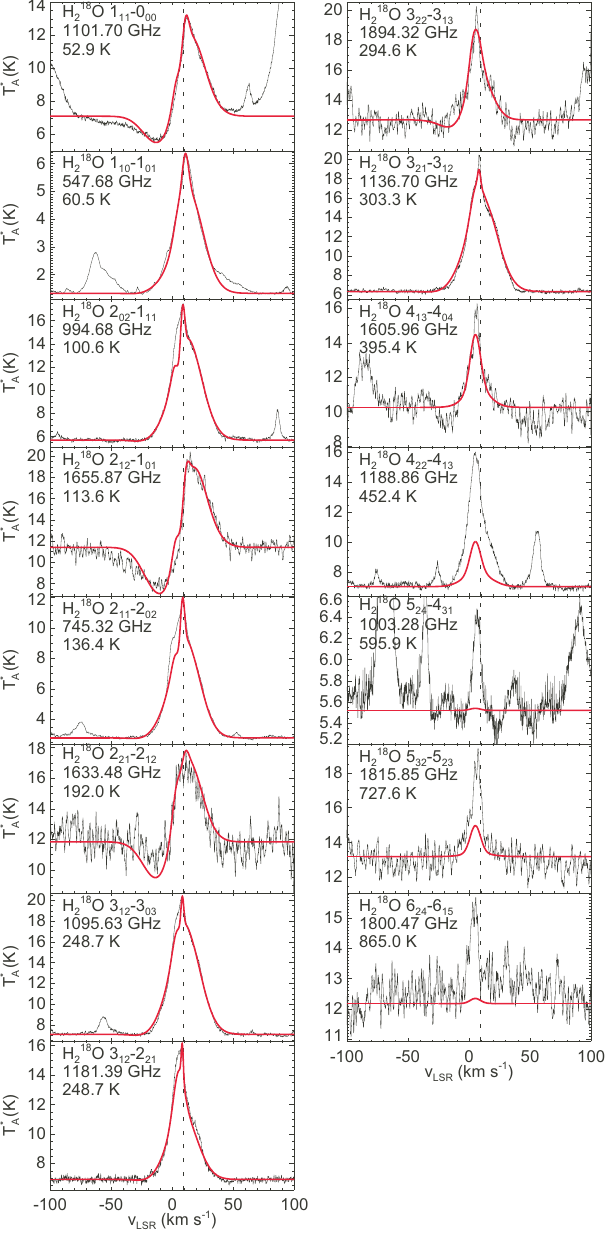}
   \caption{H$_2^{\;18}$O lines toward Orion-KL in order of increasing
   upper level energy.   
   The superposed red curves show the line profiles resulting from our radiative transfer
   modeling of the emission components and line fits to the absorption components.    
   The labels in the upper left corner of each plot list the species,
   the transition, the transition rest frequency, and the upper-level energy.  The vertical
   dashed line denotes the 9 \kms\ systemic velocity of the cloud.  The
   2$_{20\,}-\,$2$_{11}$, 4$_{23\,}-\,$3$_{30}$, 6$_{34\,}-\,$5$_{41}$, and 6$_{33\,}-\,$5$_{42}$
   spectra are omitted due to blending with other lines or a low signal-to-noise ratio. 
   We also note possible blending of the 4$_{22\,}-\,$4$_{13}$ line with
   CH$_3$OH (12$_{5\,1\,}-\,$11$_{2\,1}$) and H$_2^{\;13}$CO (18$_{2\,17\,}-\,$18$_{0\,18}$),
   both of which lie within 27\kms\ of the \iwater\ line.
   }
  \label{fig1}
\end{figure}

The physical conditions summarized in Table~2 have also been used to model the 
H$_2^{\;16}$O lines with $E_{\rm upper}\leq\:$400~K.  
To do so, the column densities of ortho- and para-H$_2^{\;16}$O are assumed to be 
500 times greater than those of \iwater, the line fluxes calculated, and then applied
using the best-fit H$_2^{\;16}$O plateau line widths determined using Eqn.~1.  For the hot core
and extended warm gas region, the H$_2^{\;16}$O widths were assumed to be twice those of
the \iwater, and the absorption components are unchanged.  The results of this simple 
approach are shown as the superposed red curves on the relevant H$_2^{\;16}$O spectra
in Fig.~2.  The potential for a more careful analysis of the H$_2^{\;16}$O and
\iiwater\ lines is illustrated by how well the constrained fits match the other line profiles,
shown as the superposed brown curves in Fig.~2.  A more detailed model 
will be presented in a future paper.

\section{Discussion}
\label{sec:disc}

Modeling of the rich spectrum of \iwater\ lines toward Orion-KL reveals several things.
First, the relatively high \water\ abundance associated with the plateau is consistent
with elevated water abundances measured previously toward KL (cf. Cernicharo \etal~2006)
as well as toward a number of other molecular outflows
(cf. Franklin \etal~2008).  This is most likely the result of a 
combination of \water-ice sublimated and sputtered from grain surfaces and \water\ formed
efficiently in the gas phase via neutral-neutral reactions favored in hotter portions of the
plateau.  The inferred water abundance for the plateau given in Table~2 is less than that
cited in some larger-beamsize studies (e.g., Harwit \etal~1998, Melnick \etal~2000b), and may
be due to the exclusion of more extended regions where 
the outflows encounter the surrounding quiescent material (cf. Genzel \& Stutzki 1989).  
These shock-heated regions, which are particularly prominent in H$_2$ emission, 
can subject virtually all of the affected gas to temperatures in excess 1000~K, thus
facilitating the neutral-neutral reactions that efficiently 
produce \water.

\begin{figure*}[!t]
   \centering
   \includegraphics[angle=0,width=1.79\columnwidth]{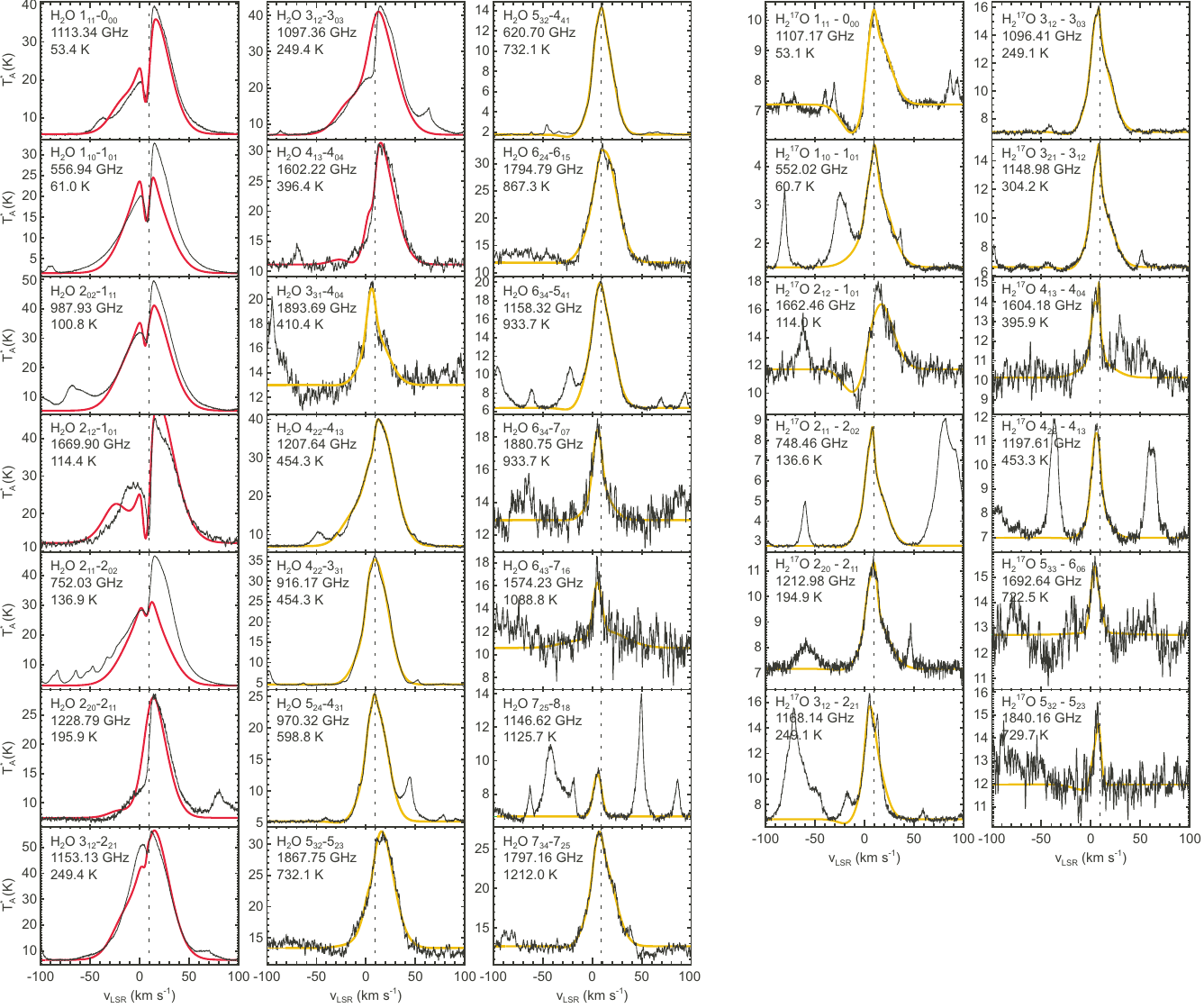}
   \vspace{-0.5mm}
   \caption{{\em Left:} Same as Fig.~1, except showing the H$_2^{\;16}$O spectra toward 
   Orion-KL/Hot Core in order of increasing upper level energy.  The red curves superposed on
   the H$_2^{\;16}$O spectra with upper-level energies less than 400~K result from our
   radiative transfer model of the emission components and line fits to the absorption components.
   The brown curves show the 
   best 5-component fit resulting from the procedure described in Section~3.
   The 2$_{21\,}-\,$2$_{12}$, 3$_{21\,}-\,$3$_{12}$, 6$_{24\,}-\,$7$_{17}$,
   7$_{34\,}-\,$7$_{25}$, and 7$_{44\,}-\,$6$_{51}$
   spectra have been omitted due to blending with other lines or a low signal-to-noise ratio.
   {\em Right:} H$_2^{\;17}$O spectra.  The brown curves superposed on the spectra show the 
   best 5-component fit resulting from the procedure described in Section~3.
   The 2$_{02\,}-\,$1$_{11}$ and 2$_{21\,}-\,$2$_{12}$
   spectra have been omitted due to blending with other lines or a low signal-to-noise ratio.
   }
  \label{fig2}
\end{figure*}

Second, the water abundances inferred for the hot core and extended warm gas are more than
an order of magnitude greater than that inferred toward other quiescent regions
(cf. Melnick \& Bergin 2005).
This is likely the result of enhanced sublimation of water-ice
from, and reduced freeze-out onto, the warm dust grains present within both regions.
It should be noted that the gas and dust temperatures inferred for the extended warm gas
should be viewed as lower limits given the probable 
presence of both water-line and continuum emission in the reference beam.

Finally, the \iwater\ ortho-to-para ratio inferred for all three emission components
is consistent with a ratio of 3:1.  A ratio of greater than 3:1 is likely the
consequence of the rather simple model adopted for each component or
residual inaccuracies in the water collisional rate coefficients, or both.

\begin{acknowledgements}
HIFI has been designed and built by a consortium of institutes and university departments from across 
Europe, Canada and the United States under the leadership of SRON Netherlands Institute for Space
Research, Groningen, The Netherlands and with major contributions from Germany, France and the US. 
Consortium members are: Canada: CSA, U.Waterloo; France: CESR, LAB, LERMA,  IRAM; Germany: 
KOSMA, MPIfR, MPS; Ireland, NUI Maynooth; Italy: ASI, IFSI-INAF, Osservatorio Astrofisico di Arcetri- 
INAF; Netherlands: SRON, TUD; Poland: CAMK, CBK; Spain: Observatorio Astron—mico Nacional (IGN), 
Centro de Astrobiolog'a (CSIC-INTA). Sweden:  Chalmers University of Technology - MC2, RSS \& GARD; 
Onsala Space Observatory; Swedish National Space Board, Stockholm University - Stockholm Observatory; 
Switzerland: ETH Zurich, FHNW; USA: Caltech, JPL, NHSC.
Support for this work was provided by NASA through an award issued by JPL/Caltech.
CSO is supported by the NSF, award AST-0540882.
\end{acknowledgements}



\pagebreak


\appendix

\begin{table*}[htdp]
\centerline{\large\bf Appendix}
\vspace{7mm}
\centerline{Table 3. Best-Fit \iwater\ Integrated Line Intensities\raisebox{0.75mm}{$\,\dagger$}}
\vspace{1.0mm}
\begin{center}
\begin{tabular}{lccccc} \hline
\rule{0mm}{5mm} & & & \multicolumn{3}{c}{Emission Component} \\*[0.2mm] \cline{4-6}
\rule{0mm}{5mm} &  &  &  &  & Extended \\
 &  & ~~Upper-Level~~ & Plateau & Hot Core & Warm Gas \\*[0.5mm]
 ~~Transition\hspace{11mm} & ~~Frequency~~ & Energy & $\int T_A^{\;*} dv$ & $\int T_A^{\;*} dv$ & 
 $\int T_A^{\;*} dv$ \\*[0.7mm]
  & (GHz) & (K) & ~~~(K-km s$^{-1}$)~~~ & ~~~(K-km s$^{-1}$)~~~ & ~~~(K-km s$^{-1}$)~~~ \\*[1mm] \hline
\rule{0mm}{5mm}\tr{1}{1}{1}{0}{0}{0} \dotfill\  & 1101.70 & ~~52.9 & 315.4 & 7.6 & 50.8 \\*[1.1mm]
\tr{1}{1}{0}{1}{0}{1} \dotfill\  & ~~547.68 & ~~60.5 & 199.0 & 8.2 & 34.6 \\*[1.1mm]
\tr{2}{0}{2}{1}{1}{1} \dotfill\ & ~~994.68 & 100.6 & 423.3 & 60.1~~ & 10.7 \\*[1.1mm]
\tr{2}{1}{2}{1}{0}{1} \dotfill & 1655.87 & 113.6 & 574.9 & 0.0 & 71.5 \\*[1.1mm]
\tr{2}{1}{1}{2}{0}{2} \dotfill & ~~745.32 & 136.4 & 276.6 & 66.3~~ & 17.7 \\*[1.1mm]
\tr{2}{2}{1}{2}{1}{2} \dotfill & 1633.48 & 192.0 & 220.6 & 10.1~~ & ~~6.4 \\*[1.1mm]
\tr{3}{1}{2}{3}{0}{3} \dotfill & 1095.63 & 248.7 & 444.9 & 70.2~~ & ~~5.2 \\*[1.1mm]
\tr{3}{1}{2}{2}{2}{1} \dotfill & 1181.39 & 248.7 & 178.8 & 50.4~~ & ~~1.8 \\*[1.1mm]
\tr{3}{2}{2}{3}{1}{3} \dotfill & 1894.32 & 294.6 & ~~65.7 & 34.7~~ & ~~2.4 \\*[1.1mm]
\tr{3}{2}{1}{3}{1}{2} \dotfill & 1136.70 & 303.3 & 424.4 & 65.3~~ & 11.8 \\*[1.1mm]
\tr{4}{1}{3}{4}{0}{4} \dotfill & 1605.96 & 395.4 & ~~46.6 & 38.6~~ & ~~0.0 \\*[1.1mm]
\tr{4}{2}{2}{4}{1}{3}\raisebox{0.55mm}{$\dagger\dagger$} \dotfill & 1188.86 & 452.4 & 
   118.5 & 57.5~~ & ~~0.2 \\*[1.1mm]
\tr{5}{2}{4}{4}{3}{1} \dotfill & 1003.28 & 595.9 & ~~~~0.0 & 6.9 & ~~2.0 \\*[1.1mm]
\tr{5}{3}{2}{5}{2}{3} \dotfill & 1815.85 & 727.6 & ~~22.2 & 32.8~~ & 10.0 \\*[1.1mm]
\tr{6}{2}{4}{6}{1}{5} \dotfill & 1800.47 & 865.0 & ~~41.5 & 36.9~~ & ~~0.0 \\*[1.2mm] \hline
\end{tabular}
\end{center}
\vspace{-0.1mm}
\hspace{23.3mm}\raisebox{0.55mm}{$\dagger$}~~The line fitting procedure is described in Section~3.\\
\rule{0mm}{4mm}
\hspace{21mm}\raisebox{0.55mm}{$\dagger\dagger$}~~Some flux attributed to this transition 
may be due to the CH$_3$OH (12$_{5\,1\,}-\,$11$_{2\,1}$) and 
H$_2^{\;13}$CO (18$_{2\,17\,}-\,$18$_{0\,18}$)\\
\phantom{0}\hspace{24.4mm} transitions, both of which lie within 27\kms\ of the \iwater\ 
\tr{4}{2}{2}{4}{1}{3} line.
\label{default}
\end{table*}%


\begin{thebibliography}{}

\bibitem[Bergin et al. 2010]{Bergin10} Bergin, E.A., Phillips, T.G., Comito, C. \ et al.\, this volume

\bibitem[Beuther et al. 2010]{Beuther10} Beuther, H., Linz, H., Bik., A., Goto, M., 
\& Henning, T. 2010, A\&A, 512, A29

\bibitem[Blake et al. 1987]{Blake87} Blake, G.~A., Sutton, E.~C., Masson, C.R., \&
Phillips, T.~G. 1987, \apj, 315, 621

\bibitem[Cernicharo et al. 2006]{Cernicharo06} Cernicharo, J., Goicoechea, J.R., Daniel, F. 
et al. 2006, \apj, 649, L33

\bibitem[Genzel \& Stutzki 1989]{Genzel89} Genzel, R. \& Stutzki, J. 1989, ARA\&A, 27, 41

\bibitem[de Graauw et al.(2010)]{deGraauw10} de Graauw, Th., Helmich, F.P., Phillips, T.G. 
et al., 2010, A\&A, 518, L6

\bibitem[Faure et al.(2007)]{Faure07} Faure, A., Crimier, N., Ceccarelli, C.,  et al. 2007, 
A\&A, 472, 1029

\bibitem[Franklin et al. 2008]{Franklin08} Franklin, J., Snell, R.L., Kaufman, M.J. 
et al. 2008, \apj, 674, 1015 

\bibitem[Gupta et al. 2010]{Gupta10} Gupta, H., Rimmer, P., Pearson, J.C. et al., this volume

\bibitem[Harwit et al. 1998]{Harwit98} Harwit, M., Neufeld, D.~A., Melnick, G.~J., \&
Kaufman, M.~J. 1998, \apj, 497, L105

\bibitem[Lerate et al. 2006]{Lerate06} Lerate, M. R., Barlow, M.J., Swinyard, B.M. 
et al. 2006, \mnras, 370, 597

\bibitem[Maret et al. 2010]{Maret10}
Maret, S., Hily-Blant, P., Pety, J., Bardeau, S., and Reynier, E. , 
in prep. (2010)

\bibitem[Melnick et al. 2000a]{Melnick00a} Melnick, G.~J., Stauffer, J.R., Ashby, M.L.N.
et al. 2000a, \apj, 539, L77

\bibitem[Melnick et al. 2000b]{Melnick00b} Melnick, G.~J., Ashby, M.L.N., Plume, R.
et al. 2000b, \apj, 539, L87

\bibitem[Melnick \& Bergin(2005)]{Melnick05} Melnick, G.~J. \& Bergin, E.~A. 2005, 
{\em Advances in Space Research}, 36, 1027

\bibitem[Phillips et al. 2010]{Phillips10} Phillips, T.G., Bergin, E.A., Lis, D.C. et al. 2010, A\&A, 518, L109

\bibitem[Pickett et al. 1998]{Pickett98}
Pickett, H.~M., Poynter, R.~L., Cohen, E.~A., et al.
1998, J. Quant.  Spectrosc. Radiat. Transfer, 60, 883

\bibitem[Pilbratt et al. (2010)]{Pilbratt10} Pilbratt, G.L., Riedinger, J.R., Passvogel, T. \ et al.\ 2010, 
A\&A, L1

\bibitem[Snell et al. 2000]{Snell00} Snell, R.~L., Howe, J.E., Ashby, M.L.N.
et al. 2000, \apjl, 539, L93 

\bibitem[Ungerechts et al. 1997]{Ungerechts97} Ungerechts, H., Bergin, E.A., Goldsmith, P.F.
et al. 1997, \apj, 482, 245 

\bibitem[Wright et al. 2000]{Wright00} Wright, C.M., van Dishoeck, E.F., Black, J.H. 
et al. 2000, A\&A, 358, 689

\end{thebibliography}
\end{document}